\documentstyle[aps,amsfonts,epsfig]{revtex}
\DeclareMathSymbol{\restriction}{\mathrel}{AMSa}{"16}
\newcommand{\Tr}{{\rm Tr\,}}

\newcommand{\mmix}{\bbox{\,\succ\,}}
\newcommand{\hs}{h_{\rm s}}
\newcommand{\hsd}{{h_{\rm s}}'}
\newcommand{\llabel}[1]{{\bf \underline{#1}}\qquad\quad\label{#1}}
\renewcommand{\llabel}[1]{\label{#1}}
\newcommand{\bib}[1]{\bibitem{#1}}
\begin{document}
\draft
\title{On the Curvature of Monotone  Metrics and a  \\Conjecture
Concerning  the Kubo-Mori Metric}
\author{J.~Dittmann\thanks{e-mail
correspondence: {\tt
dittmann@mathematik.uni-leipzig.de}}}\address{ Leipzig, Germany}
\date{May 30, 1999}
\maketitle
\begin{abstract}
Monotone metrics on the space of positive matrices generalize the
classical Fisher metric  of statistical distinguishability
of probability distributions to the quantum case. These
metrics are in
one to one correspondence with operator monotone functions. It is the aim
of this article to determine curvature quantities of  an
arbitrary Riemannian monotone metric using the Riesz-Dunford operator
calculus as the main technical tool. The resulting scalar curvature is
explained in more detail for three examples.
In particular, we show an important
conjecture of Petz concerning the Kubo-Mori metric up to a formal proof of
the concavity of a certain function on ${\Bbb R}_+^3$. This
concavity  seems to be
numerically evident. The conjecture asserts that the scalar curvature of
the Kubo-Mori metric increases if one goes to more mixed states.
\end{abstract}\pacs{}
\section{Introduction}
Let ${\cal D}_n$ resp.~${\cal D}_n^1$ denote the
manifold of complex
positive $n\times n$-matrices (of trace one).
A Riemannian metric on ${\cal D}_n^1$ is called monotone if it is
decreasing under stochastic mappings (the exact definition is given below).
These metrics are of  interest in quantum statistics and
information theory since ${\cal D}_n^1$ is
the space of (faithful) states (represented by positive density matrices)
of an $n$-level quantum system. Monotone metrics  generalize  the Fisher
information (metric) of  classical statistics,\cite{Amari,Friedrich}, to
the quantum case. Indeed, the restrictions of any
monotone metric to the submanifold of  diagonal  matrices
(probability distributions on an $n$-point set)
equals (up to a factor) the Fisher metric.
Roughly speaking, the Fisher metric  measures the statistical
distinguishability of probability distributions and a similar meaning
is expected in the quantum case, see
e.g.~\cite{Braunstein,Braunstein96,Twamley,Paraoanu}.
Monotone metrics can be seen as the Hessian of  entropy-like
functions.  The relation between entropies, monotone metrics and operator
monotone functions has been considered in several articles ,
see \cite{Rus} and
references found there. Moreover, in \cite{UD} it has been  explained how
monotone metrics appear  on the background of the concept of
purification of mixed states \cite{Uh91}. There monotone metrics  arise
from the reduction of certain metrics on the space of pure states of a
larger quantum system.
The underlying geometry is that of a Riemannian submersion, \cite{Besse},
where the  metric in the purifying space is built by the
natural modular operator of the Tomita-Takesaki theory.

A  monotone metric of special interest for mathematical and
physical reasons, \cite{Braunstein,Twamley,Paraoanu,Di98,UD}, is
the so called Bures metric introduced by Uhlmann, \cite{Uh92b}.
Its curvature quantities are known, \cite{Di93a,Di99}. Moreover,
certain curvature results were obtained for the Kubo-Mori metric
in \cite{Petz94}. It is the first aim of this paper to obtain
similar results for the general case (Propositions 1-4). These
considerations are based  on a classification result of Petz,
\cite{Petz}. Continuing the work of Morozova and Chentsov,
\cite{Moro}, who initiated the study of monotone metrics, he
proved a one to one correspondence between monotone metrics and
operator monotone functions (cf.~\cite{Donog,Kubo}). Thus, a
monotone metric is of the form $$g_\varrho(X,Y)=\Tr X^* c({\bf
L}_\varrho,{\bf R}_\varrho)(Y)\,. $$ For calculating  curvature
quantities we have  to handle derivatives with respect to
$\varrho$ of the function $c$ of
the operators of left and right multiplication.
For this purpose we use the Riesz-Dunford operator
calculus and obtain the Riemannian curvature tensor for an
arbitrary monotone metric. Several authors called objects similar
to $c({\bf L},{\bf R})$ superoperators  due to  the dependence on
$\varrho$. We   regard $c({\bf L},{\bf R})$ as a field of
operators acting on a bundle in order to reflect the underlying
geometry.

Special interest we focus on the scalar curvature. The main
technical difficulty here is due to   the large number of terms
one meets. However, we end up with a suitable expression (Theorem
1) for the scalar curvature  and apply  this result to three
examples, the Bures, the largest and the Kubo-Mori metric (Theorem
2). In the last case there is a conjecture
of Petz,\cite{Petz94}, that the
scalar curvature increases if one goes to more mixed states. It is
the second aim of this paper to reduce the proof of this important
conjecture to the concavity of a certain function on ${\Bbb
R}_+^3$ (Theorem 3). This concavity  seems to be numerically
evident by many experiences and several function plots, although
we do not have a formal proof up to now. Since the function in
question, $\hs$, see formula (\ref{ds}), is homogeneous
 the asserted concavity is, actually, a certain property of
the function  $\hs(x,y,1)$ in two variables. Of course,
this property is stronger than concavity.

The paper is organized as follows. In Section II
we fix the framework of the Riesz-Dunford calculus for our purposes.
Section III gives the basic definition and classification theorem for
monotone metrics and contains some preliminary remarks on
Morozova-Chentsov functions. In Sections IV and V we determine the
curvature quantities including the examples. Finally, Section VI
deals with the monotonicity conjecture.

\noindent
{\bf Notations:}
Let  ${\cal M}_n({\Bbb C})$  denote the space of complex $n{\times}
n$-matrices and ${\cal D}_n$ (resp.~${\cal D}_n^1$) the manifold of positive
definite complex matrices $\varrho$ (of trace one).
${\cal D}_n^{(1)}$ is an open subset of the space of Hermitian matrices
(of trace one). Therefore, we identify
the tangent spaces
${\rm T}_\varrho{\cal D}_n^{(1)}$ of these manifolds  in an
obvious way with the  space of  Hermitian (traceless for
${\cal D}_n^1$) matrices   and consider vector
fields $\cal X,Y,\dots$
as functions on ${\cal D}_n^{(1)}$ with values in these vector spaces.
By the corresponding italic symbols we mean vectors tangent at
$\varrho$ resp.~the values of a vector field  at $\varrho$,
$X={\cal X}_\varrho$, etc.~for the fixed point $\varrho$ we have in mind.
$[{\cal X,Y}]$ is understood as
the commutator of these matrix valued functions, i.e.
$[{\cal X,Y]_\varrho:=
\cal X_\varrho Y_\varrho-\cal X_\varrho Y_\varrho}=[X,Y]$,
whereas the usual commutator of vector fields is denoted by
$[{\cal X,Y}]_{\rm vf}$.
By ${\bf L}_\varrho$ and ${\bf R}_\varrho$ we denote the operators
of left resp.~right multiplication by $\varrho\in{\cal D}_n^{(1)}$.
Therefore, ${\bf L}$ and ${\bf R}$ are fields of operators.
But, similarly to
other geometrical object, we frequently omit the index $\varrho$ for
simplicity of notation even if we have a concrete point $\varrho$ in mind.
Moreover, if $\bf L$ or $\bf R$ are indicated otherwise we  mean the
corresponding multiplication operators.
\section{Operator Calculus}
Let $f$ be  a complex analytic function defined on a neighborhood
of ${\Bbb R}_+$. Then we have by the Riesz-Dunford operator calculus
$$
f(\varrho)=
\frac{1}{2\pi{\bf i}}\oint f(\xi)
(\xi-\varrho)^{-1}\,{\rm d}\xi\,,
$$
where $\xi(t)$ is a path surrounding the positive spectrum of $\varrho$.
For a diagonal $\varrho$ this is just  Cauchy' integral formula
applied to each eigenvalue. We notice, that
the operators ${\bf L}_\varrho$ and ${\bf R}_\varrho$ are selfadjoint
w.r.~to the Hermitian form $X,Y\mapsto {\rm Tr\,X^*\,Y}$
on ${\cal M}_n({\Bbb C})$ and their common
spectrum equals the   spectrum of $\varrho$ (with multiplicity $n$).
Thus, the function $f$ of the operator ${\bf L}_\varrho$,
$f\left({\bf L}_\varrho\right)$, is well defined
and has, again by the Riesz-Dunford calculus, the
representation
\begin{equation}\llabel{int1}
f\left({\bf L}_\varrho\right)=
\frac{1}{2\pi{\bf i}}\oint f(\xi)
\left(\xi{\bf Id}-{\bf L}_\varrho\right)^{-1}\,{\rm d}\xi\,.
\end{equation}

Since left and right
multiplication commute a similar reasoning applies to a function $c$,
complex analytic in two variables, of ${\bf L}_\varrho$ and
${\bf R}_\varrho$.
This results in
\begin{equation}\llabel{int2}
c({\bf L}_\varrho,{\bf R}_\varrho)=
\frac{1}{(2\pi{\bf i})^2}\oint\!\!\oint c(\xi,\eta)
\frac{1}{\xi{\bf Id}-{\bf L}_\varrho}\circ
\frac{1}{\eta{\bf Id}-{\bf R}_\varrho}
\,{\rm d}\xi\,{\rm d}\eta\,
\end{equation}
or, equivalently,
\begin{equation}\llabel{int2a}
c({\bf L}_\varrho,{\bf R}_\varrho)(Y)=
\frac{1}{(2\pi{\bf i})^2}\oint\!\!\oint c(\xi,\eta)
\frac{1}{\xi-\varrho}Y
\frac{1}{\eta-\varrho}
\,{\rm d}\xi\,{\rm d}\eta\,,
\end{equation}
where we integrate in every variable once around the  spectrum of
$\varrho$.
These integral formulae are the main technical tool for what follows,
because they allow for determining derivatives of  functions of
$\varrho$, $\bf L$ and  $\bf R$, e.g.~
\begin{eqnarray*}
\frac{\rm d}{{\rm d}t}f\left(\varrho+t X\right)_{|t=0}&=&
\frac{1}{2\pi{\bf i}}\oint f(\xi)
\frac{1}{\xi-\varrho}\, X\,
\frac{1}{\xi-\varrho}
\,{\rm d}\xi\\
\frac{\rm d}{{\rm d}t}f\left({\bf L}_{\varrho+t X}\right)_{|t=0}&=&
\frac{1}{2\pi{\bf i}}\oint f(\xi)
\frac{1}{\xi{\bf Id}-{\bf L}_{\varrho}}\circ
{\bf L}_X\circ
\frac{1}{\xi{\bf Id}-{\bf L}_{\varrho}}
\,{\rm d}\xi\,,\qquad
X=X^*\in{\rm T}_\varrho {\cal D}_n\subset {\cal M}_n({\Bbb C}).
\end{eqnarray*}
The last expression is just the covariant derivative of $f({\bf L})$
in the direction $X$ w.r.~to the local flat affine structure on
${\cal D}$ inherited from
$\left\{M\in {\cal M}_n({\Bbb C})|M=M^*\right\}$.
We denote this flat covariant derivative by
${\rm D}$.
The derivative ${\rm D}_{\!X}c({\bf L},{\bf R})$
and the second order derivatives which we need later on
are obtained similarly
from (\ref{int2}) or  (\ref{int2a}) applying the Leibniz rule
to the composition of
resolvents.
Since these formulae are obvious, we do not write down them here.

The right hand sides   of the last two
equations become more explicit if we decompose the tangent vector
as $X=Y+[A,\varrho]$, where $[Y,\varrho]=Y\varrho-\varrho Y=0$ and
$A=-A^*$ (The existence of such a
decomposition can be seen assuming that $\varrho$ is diagonal). Now a
simple calculation yields
\begin{eqnarray*}
&&
\frac{\rm d}{{\rm d}t}f\left(\varrho+t X\right)_{|t=0}=
Y f'(\varrho)+[A,f(\varrho)]\\
{\rm D}_{\!X}f({\bf L})_\varrho&=&
\frac{\rm d}{{\rm d}t}f\left({\bf L}_{\varrho+t X}\right)_{|t=0}=
{\bf L}_Y\circ f'({\bf L}_\varrho)+
\left[{\bf L}_A,{\bf L}_{f(\varrho)}\right]=
{\bf L}_{Y f'(\varrho)+[A,f(\varrho)]}\,.
\end{eqnarray*}
But, in order to  remove
integrals in the forthcoming second order derivatives
we would have to consider more subtle decompositions.
Moreover, the functions $c$ we regard are of the
form $1/(f(s/t)t)$, so that, actually,
one  needs only one integration to  consider
$f(\bf L_\varrho/\bf R_\varrho)$.
In \cite{DiRu} a  calculus  for its first  order
derivatives has been  developed, which leads for diagonal $\varrho$,
essentially, to a Pick matrix method found in
\cite{Donog}, Chap.~VIII.
However, here we need the second order derivatives as well, and we will
not try  to avoid the integrals. The approach used here has the advantage
of symmetry.
\section{Monotone Metrics}
 A family of Riemannian metrics $g=\left\{g^n\right\}$
on $\left\{{\cal D}_n^1\right\}$, ${n\in{\Bbb N}} $,
is called monotone iff
$$g^{m}_{{\bf T}(\varrho)}({\bf T}_*(X),{\bf T}_*(X))\leq
g^{n}_\varrho(X,X)
$$
holds for every stochastic mapping
${\bf T}: {\cal M}_n({\Bbb C}) \rightarrow {\cal M}_m({\Bbb C}) $ and all
$\varrho\in{\cal D}_n^1$,  $X\in{\rm T}_\varrho {\cal D}_n^1$.
For more details we
refer to \cite{Petz}, where Petz considered, roughly speaking,
monotone Hermitian forms  on the complexified tangent bundles of
${\cal D}_n^1$.
The above defined monotone Riemannian metrics  in \cite{Petz} are
called
symmetric monotone metrics and the  real part of  a monotone Hermitian
form is, of course,  symmetric.
We will use a slightly different terminology.
By a monotone metric we will always mean  a
monotone Riemannian metrics, since we are interested here only
in those ones.
From now on we fix an integer  $n\geq 2$  and  omit the index $n$
at $g^n$ and ${\cal D}_n^{(1)}$.
Thus, we mean by $g$ either the whole family
or the concrete metric related to the number $n$ we have in mind.

\vspace{0.3cm}
\noindent
{\bf Theorem \cite{Petz}:} {\sl There is a one to one
correspondence between monotone Riemannian metrics $g$  and
operator monotone
functions $f: {\Bbb R}_+\rightarrow{\Bbb R}_+$ satisfying
the symmetry condition
$f(t)=t\,f(1/t)$.
This  correspondence is given by}
$$
g_\varrho(X,Y)={\rm Tr}\,X\frac{1}{f({\bf
L}_\varrho /{\bf R}_\varrho ){\bf R}_\varrho }(Y)\,,
\qquad X,Y\in{\rm T}_\varrho {\cal D}^1.
$$

\vspace{0.3cm}
\noindent
For the definition, integral representation and main properties of
operator monotone functions we refer to \cite{Donog,Kubo}.
A crucial point for what follows is the fact, that such a function has a
complex analytic extension to the open right complex half plane such that
$f(\overline{x})=\overline{f(x)}$. Thus, the
function c
defined by\begin{equation}
c(x,y):=\frac{1}{f(x/y)y}\,,
\end{equation}
called the Morozova-Chentsov
function related to the monotone metric $g$, is analytic in both arguments
on an open neighborhood of ${\Bbb R}_+^{\;2}\subset{\Bbb C}^2$ and
the metric $g$ takes the more appropriate  for our
purposes form
\begin{equation}\llabel{metric1}
g({\cal X},{\cal Y})={\rm Tr}\,{\cal X}\,c({\bf L},{\bf R})({\cal Y})\,,
\end{equation}
or, equivalently,
\begin{equation}\llabel{metric2}
g_{\varrho}(X,Y)= \frac{1}{(2\pi{\bf i})^2}\,{\rm Tr}\oint\!\!\oint
c(\xi,\eta)X(\xi-\varrho)^{-1}Y(\eta-\varrho)^{-1}\, \,{\rm d}\xi\,{\rm
d}\eta\,.
\end{equation}
Clearly, the above
formulae also define a Riemannian metric on ${\cal D}$, which we also
denote by $g$. In the next sections
we will first deal with this manifold and then consider ${\cal D}^1$
as a Riemannian submanifold of codimension 1.

Some examples of symmetric operator monotone functions are
$$f(x)=
\frac{x+1}{2}\,,\quad \frac{2x}{x+1}\,,\quad\frac{x-1}{\ln x}$$
leading to the Morozova-Chentsov functions
\begin{equation}\llabel{MCf}
c(x,y)=
\frac{2}{x+y}\,,\quad \frac{x+y}{2xy}\,,\quad\frac{\ln x-\ln y}{x-y}
\end{equation}
and to the smallest, largest and Kubo-Mori metric. For further examples
we refer to \cite{Kubo,Petz}.  For the time being we are interested
in the general case and return to these examples later.

The  calculations of the following sections make use of certain
algebraic and differential properties common for all Morozova-Chentsov
functions. Multiplying a monotone  metric by a factor if necessary,
we will assume from now on that the corresponding
operator monotone function $f$ is normalized by
$f(1)=1$. This and the above mentioned properties of $f$ imply
\begin{equation}\llabel{c0}
c(x,x)=\frac{1}{x}
\end{equation}
\begin{equation}\llabel{c1}
c(\overline{x},\overline{y})=\overline{c(x,y)}\,,\qquad
c(x,y)=c(y,x)\,,\qquad
c(x,y)=t\;c(t \,x,t\, y)\,,\quad t\in{\Bbb R}_+\,.
\end{equation}
Lets  denote by  $c^{(k,l)}$ the partial derivatives of $c$
of order $(k,l)$. Then we have by the symmetry of $c$
\begin{equation}\llabel{c2}
c^{(k,l)}(x,y)=c^{(l,k)}(y,x)
\end{equation}
and differentiating the last equation of (\ref{c1}) w.r.~to $t$ and
$x,y$ yields  the further identities
\begin{eqnarray}
c(x,y)+x\,c^{(1,0)}(x,y)+y\,c^{(0,1)}(x,y)&=&0\nonumber\\
\llabel{c3}
2\,c^{(1,0)}(x,y)+y\,c^{(1,1)}(x,y)+x\,c^{(2,0)}(x,y)&=&0\\
2\,c^{(0,1)}(x,y)+x\,c^{(1,1)}(x,y)+y\,c^{(0,2)}(x,y)&=&0\,.\nonumber
\end{eqnarray}
In particular,
\begin{equation}\llabel{c4}
c^{(1,0)}(x,x)=c^{(0,1)}(x,x)=-\frac{1}{2x^2},
\end{equation}
but there is no such identity for  $c^{(2,0)}(x,x)$, cf.~(\ref{MCf}).

We will have to do with partial derivatives  of order at most two. Thus we
can always remove the mixed derivatives $c^{(1,1)}$ and replace
$c^{(0,k)}$ by $c^{(k,0)}$ changing the order of arguments.
For simplicity we write $c'$ and $c''$ instead of
$c^{(1,0)},c^{(2,0)}$.
Exhausting the above relations we can attain that $c^{(k,0)}$, $k\leq2$,
has   arguments ordered alphabetically. This uniqueness is useful for
comparing expressions,  however, we will
not always insist on such an ordering, since it may destroy the symmetry.
Therefore, we leave it at the reduced equations  (\ref{c3})
\begin{eqnarray}
c(x,y)+x\,c'(x,y)+y\,c'(y,x)&=&0\,\nonumber\\
\llabel{c5}
2x\,c'(x,y)+x^2c''(x,y)&=&
2y\,c'(y,x)+y^2c''(y,x)\,.
\end{eqnarray}
Finally, it will be convenient to use  partial logarithmic derivatives
instead of usual ones and we apply the above notation to the function
$\ln c$ too,
\begin{equation}\llabel{c6}
(\ln c)'(x,y):=\frac{\partial \ln c (x,y)}{\partial x}=
\frac{c'(x,y)}{c(x,y)}\,.
\end{equation}
For example, the first identity of (\ref{c3}) now reads
$$1+x\, (\ln c)'(x,y)+y\, (\ln c)'(y,x)=0\,.$$
\section{Covariant Derivative and Curvature Tensor}
\subsection{Not Normalized Case}
As already mentioned in Section II the manifold $\cal D$ carries
locally a natural
flat affine structure related to the usual parallel
displacement. The
corresponding covariant derivative ${\rm D}$ is the derivation
along straight lines, i.e.~
\begin{equation}
({\rm D}_{\!Z}{\cal T})_\varrho=
\frac{{\rm d}}{{\rm d}t}\,{{\cal T}_{\varrho+t \,Z}}_{\restriction t=0}
\,,\qquad Z\in{\rm T}_\varrho{\cal D}\,,
\end{equation}
for any tensor field $\cal T$ on $\cal D$.
In particular
\begin{equation}
\llabel{DN}
{\rm D }_Z{\varrho}=Z\,.
\end{equation}
Since the trace and the matrix
multiplication are compatible with this flat affine structure
the Leibniz rule implies
\begin{equation}
{\rm D}_{\!Z}(\xi-\varrho)^{-1}=
(\xi-\varrho)^{-1}Z\,(\xi-\varrho)^{-1}\,,\qquad\xi\in{\Bbb C}\,.
\end{equation}
From (\ref{int2a}) we obtain for  the flat covariant derivative of
$c({\bf L},{\bf R})$
\begin{equation}\llabel{nc(L,F)}
{\rm D}_{\!Z} c({\bf L},{\bf R})_\varrho\;(Y)=
\frac{1}{(2\pi{\bf i})^2}\oint\!\!\oint c(\xi,\eta)
\left\{
\frac{1}{\xi-\varrho}Z\frac{1}{\xi-\varrho}Y\frac{1}{\eta-\varrho}+
\frac{1}{\xi-\varrho}Y\frac{1}{\eta-\varrho}Z\frac{1}{\eta-\varrho}
\right\}
\,{\rm d}\xi\,{\rm d}\eta\,,
\end{equation}
and, therefore, using (\ref{metric1})
\begin{equation}\llabel{nmetric}
{\rm D}_{\!Z}g\;(X,Y)=
\frac{1}{(2\pi{\bf i})^2}\,{\rm Tr}\oint\!\!\oint c(\xi,\eta)\,Z\,\left\{
\frac{1}{\xi-\varrho}\,X\,\frac{1}{\eta-\varrho}\,Y\,\frac{1}{\xi-\varrho}+
\frac{1}{\xi-\varrho}\,Y\,\frac{1}{\eta-\varrho}\,X\,\frac{1}{\xi-\varrho}
\right\}\,{\rm d}\xi\,{\rm d}\eta\,.
\end{equation}
Similarly we proceed we the flat derivative of second order
with respect to vector fields $\cal X$ and  $\cal Y$.
We denote it by  ${{\rm D}}^2_{\!\cal X,Y}$,  i.e.
$${{\rm D}}^2_{\!\cal X,Y}:=
{\rm D}_{\!\cal X}{\rm D}_{\cal Y}-
{\rm D}_{{\rm D}_{\!\cal X }\cal Y}.$$
Clearly the differential operator ${{\rm D}}^2_{\!\cal X,Y}$ at $\varrho$
depends only on the values of the fields at this
point, cf.~\cite{Besse}.

From (\ref{nmetric}) we find using the Leibniz rule
\begin{eqnarray}
{{\rm D}}^2_{\!Z,W}\!g\;(X,Y)&=&
\frac{1}{(2\pi{\bf i})^2}\,{\rm Tr}\oint\!\!\oint c(\xi,\eta)\,\left\{
\left(
Z\frac{1}{\xi-\varrho}W+
W\frac{1}{\xi-\varrho}Z\right)
\frac{1}{\xi-\varrho}
X\frac{1}{\eta-\varrho}Y
\frac{1}{\xi-\varrho}\right.\nonumber\\
\llabel{nnmetric}
&&\hspace{3.5cm}+\left.
Z\frac{1}{\xi-\varrho}X\frac{1}{\eta-\varrho}W\frac{1}{\eta-\varrho}Y
\frac{1}{\xi-\varrho}
\right\}
{\rm d}\xi\,{\rm d}\eta+\{X\leftrightarrow Y\}\,.
\end{eqnarray}
The symbol $\{X\leftrightarrow Y\}$ means the  expression before
with exchanged $X$ and $Y$.

\bigskip
Now let $\nabla$ be the covariant derivative related to the Levi-Cività
connection of  the metric $g$. It is uniquely determined by
$$
2g(\nabla_{\!\cal X}{\cal Y},{\cal Z})=
{\cal X}g({\cal Y},{\cal Z})+
{\cal Y}g({\cal X},{\cal Z})-
{\cal Z}g({\cal X},{\cal Y})+
g([{\cal X},{\cal Y}]_{\rm vf},{\cal Z})+
g([{\cal Z},{\cal X}]_{\rm vf},{\cal Y})+
g({\cal X},[{\cal Z},{\cal Y}]_{\rm vf})\,.
$$
Moreover,  $\nabla_{\!\cal X}{\cal Y}$ is of the
form
\begin{equation}\llabel{Gamma}
\nabla_{\!\cal X}{\cal Y}=
{\rm D}_{\!{\cal X}}{\cal Y}+\Gamma ({\cal X,Y})\,,
\end{equation}
where $\Gamma $ is a certain symmetric (1,2)-tensor field on
$\cal D$, because every two torsionless covariant derivatives
are related in this way.
In an affine coordinate system on $\cal D$ it would be
described by the usual Christoffel symbols $\Gamma_{i,j}^k$.
This allows us to use the symbol $\Gamma$, although, in general
the Christoffel symbols are not of tensorial type.
Now the above defining equation  turns out
to be equivalent  to the  requirement that
\begin{equation}\llabel{defequ}
2g(\Gamma ({ X,Y}),{ Z})=
{\rm D}_{\!X} g\;({ Y,Z})+
{\rm D}_{Y}g\;({ X,Z})-
{\rm D}_{\! Z}g\;({ X,Y})
\end{equation}
holds for all  tangent vectors  $X,Y,Z$. This corresponds to the
formulae for the Christoffel symbols in terms of
the coefficients  of the metric tensor and their first order
derivatives.

\bigskip
\noindent
{\bf Proposition 1:} {\sl Let $X=X^*,Y=Y^*\in{\rm T}_\varrho{\cal D}$,
then}
\begin{eqnarray}
&&\Gamma (X,Y)=\nonumber\\
&&\label{S2}
\frac{1}{2(2\pi{\bf i})^2 c({\bf L},{\bf R})}\oint\!\!\oint
c(\xi,\eta)\left\{
\frac{1}{\xi{-}\varrho}X\frac{1}{\xi{-}\varrho}Y\frac{1}{\eta{-}\varrho}+
\frac{1}{\eta{-}\varrho}X\frac{1}{\xi{-}\varrho}Y\frac{1}{\xi{-}\varrho}-
\frac{1}{\xi{-}\varrho}X\frac{1}{\eta{-}\varrho}Y\frac{1}{\xi{-}\varrho}
\right\}{\rm d}\xi{\rm d}\eta+\{X\leftrightarrow Y\}
\end{eqnarray}

\bigskip
\noindent {\bf Proof:} The right hand side of (\ref{S2}) is Hermitian
by (\ref{c1}), i.e.~it is a vector tangent to
$\cal D$. Using (\ref{nmetric})   one verifies straightforwardly
that (\ref{S2}) fulfills (\ref{defequ}), where
one can assume for brevity of calculation  that $Y=X$. \hfill$\Box$

\bigskip
Next we consider the Riemannian curvature tensor given by
\begin{equation}\llabel{RC}
{\cal R}({\cal X,Y,Z,W}):=g\left({\cal X},
\nabla_{\!\cal Z}\nabla_{\!\cal W}{\cal Y}-
\nabla_{\!\cal W}\nabla_{\!\cal Z}{\cal Y}-
\nabla_{[{\cal  Z,W}]_{\rm vf}}{\cal Y}
\right)\,.
\end{equation}
It can be expressed in terms  of $\Gamma$ and ${\rm D}^2g$,
indeed we show:

\bigskip
\noindent
{\bf Proposition 2:} {\sl
\begin{eqnarray}
{\cal R}({\cal X,Y,Z,W})&=&
g(\Gamma (X,W),\Gamma (Y,Z))-g(\Gamma (X,Z),\Gamma (Y,W))\nonumber\\
&&\llabel{Prop2}
+\frac{1}{2}\left(
{{\rm D}}^2_{\!{\cal X,W}}g\;({\cal Y,Z})
+{{\rm D}}^2_{{\cal Y,Z}}g\;({\cal X,W})
-{{\rm D}}^2_{{\cal X,Z}}g\;({\cal Y,W})
-{{\rm D}}^2_{{\cal Y,W}}g\;({\cal X,Z})
\right)\,,
\end{eqnarray}
where ${{\rm D}}^2g$ and $\Gamma$  are given
by (\ref{nnmetric})} and (\ref{S2}).

\bigskip
{\bf Proof:} Actually (\ref{Prop2}) corresponds to
the known formula for the coefficients of the curvature  tensor
in terms of Christoffel symbols and second order derivatives of the
metric coefficients w.r.~to a coordinate system.
Thus we  derive (\ref{Prop2}) from (\ref{RC}) not appealing to the
particular metric we have in mind.
In order to simplify this calculation  let us assume for the
moment that $\cal X,Y,Z$ and $\cal W$ are parallel vector fields w.r.~to
the underlying flat affine structure, i.e.~they are constant Hermitian
matrix valued functions. Then the flat derivative
${\rm D}_{\!\cal X}{\cal Y}$
and the commutator
$[{\cal X,Y}]_{\rm vf}=
{\rm D}_{\!\cal X}{\cal Y}-
{\rm D}_{\!\cal Y}{\cal X}$
vanish for all pairs of these fields
and
${{\rm D}}^2_{\!\cal X,Y}=
{\rm D}_{\!\cal X}{\rm D}_{\!\cal Y}=
{\rm D}_{\!\cal Y}{\rm D}_{\!\cal X}$.
Thus we get using $\nabla g=0$ and (\ref{Gamma})
\begin{eqnarray*}
{\cal R}({\cal X,Y,Z,W})
&=&g\left({\cal X},
\nabla_{\!\cal Z}\nabla_{\!\cal W}{\cal Y}-
\nabla_{\!\cal W}\nabla_{\!\cal Z}{\cal Y} \right)\nonumber\\
&=&
\nabla_{\!\cal Z}\left(g({\cal X},\nabla_{\!\cal W}{\cal Y})\right)
-g(\nabla_{\!\cal Z}{\cal X},\nabla_{\!\cal W}{\cal Y})
-\nabla_{\!\cal W}\left(g({\cal X},\nabla_{\!\cal Z}{\cal Y})\right)
+g(\nabla_{\!\cal W}{\cal X},\nabla_{\!\cal Z}{\cal Y})\nonumber\\
&=&g(\Gamma ({\cal X,W}),\Gamma ({\cal Y,Z}))-
   g(\Gamma ({\cal X,Z}),\Gamma ({\cal Y,W}))\nonumber\\
&&+{\cal Z}\left(g({\cal X},\Gamma ({\cal Y,W}))\right)
-  {\cal W}\left(g({\cal X},\Gamma ({\cal Y,Z}))\right)\,.
\end{eqnarray*}
But the  last two terms yield by (\ref{defequ})
\begin{eqnarray*}
&&\frac{1}{2}{\rm D}_{\!\cal Z}\left\{
 {\rm D}_{\!\cal Y}g\;({\cal X,W})
+{\rm D}_{\cal W}g\;({\cal X,Y})
-{\rm D}_{\!\cal X}g\;({\cal Y,W})\right\}
-\frac{1}{2}{\rm D}_{\cal W}\left\{
 {\rm D}_{\cal Y}g\;({\cal X,Z})
+{\rm D}_{\!\cal Z}g\;({\cal X,Y})
-{\rm D}_{\!\cal X}g\;({\cal Y,Z})
                           \right\}\\
&=&
\frac{1}{2}\left(
 {{\rm D}}^2_{\!\cal X,W}g\;({\cal Y,Z)}
+{{\rm D}}^2_{\!\cal Y,Z}g\;({\cal X,W)}
-{\rm D}^2_{\!\cal X,Z}g\;({\cal Y,W})
-{{\rm D}}^2_{\cal Y,W}g\;({\cal X,Z)}
\right)\,.
\end{eqnarray*}
{}\hfill$\Box$
\subsection{Normalized Case}
The knowledge of the covariant derivative and  the curvature  of
$({\cal D},g)$ allows for determining  these quantities for the
codimension one submanifold
$({\cal D}^1,g)$. We denote them by  ${}^1\nabla$ and ${\cal R}^1$.
For this purpose let $\cal X,Y\dots$ be vector fields  on
${\cal D}^1$ extended to some neighborhood  of
${\cal D}^1$ in ${\cal D}$.

First of all we observe that the radial vector field $\cal N$ given by
\begin{equation}\llabel{N}
{\cal N}_\varrho:=\varrho\,, \qquad \varrho\in{\cal D}\,,
\end{equation}
is perpendicular to
all submanifolds of positive matrices with fixed  trace, in particular
to ${\cal D}^1$.
Indeed, let $X\in {\rm T}_\varrho {\cal D}$ with  $\Tr X=0$, then
$$
g(X,\varrho)={\rm Tr\,}{X}\,
c({\bf L}_\varrho,{\bf R}_\varrho)(\varrho)=
{\rm Tr}\,X\,c(\varrho,\varrho)\varrho=
{\rm Tr}\,X\,\varrho^{-1}\varrho=
{\rm Tr}\,X=0\,,
$$
where we used $c(x,x)=1/x$, see (\ref{c0}).
Moreover, this field is normalized at ${\cal D}^1$,
$g({\cal N,N})_\varrho=
g_\varrho(\varrho,\varrho)=\Tr \varrho=1$ and satisfies
\begin{equation}\llabel{DXN}
{\rm D}_{\cal X}{\cal N}={\cal X}
\end{equation}
for all
vector fields $\cal X$ on $\cal D$.
Actually, this is a more precise formulation of (\ref{DN}).
The following relations concerning this radial field are
less obvious.

\bigskip\noindent
{\bf Lemma 1}: For all vector fields  $\cal X,Y,Z$ on
$\cal D$ holds
\begin{eqnarray}\llabel{l1}
{\rm i)}&\qquad&\Gamma({\cal X,N})=-\frac{1}{2}\,X\\
\llabel{l2}
{\rm ii)}&\qquad&
g(\Gamma({\cal X,Y}),{\cal N})=-\frac{1}{2}\,g({\cal X,Y})\\
\llabel{l3}
{\rm iii)}&\qquad&
{\cal R(X,Y,Z,N})=0
\end{eqnarray}

\bigskip
\noindent
{\bf Proof:}

\noindent
i) From (\ref{S2}) and (\ref{c3}) we get
\begin{eqnarray*}
\Gamma ({\cal X,N})_\varrho&=&
\frac{1}{2(2\pi{\bf i})^2 c({\bf L},{\bf R})}\oint\!\!\oint
c(\xi,\eta)\left\{
\frac{\varrho}{(\xi{-}\varrho)^2}X\frac{1}{\eta-\varrho}+
\frac{1}{\xi-\varrho}X\frac{\varrho}{(\eta-\varrho)^2}
\right\}{\rm d}\xi{\rm d}\eta\\
&=&
\frac{1}{2\, c({\bf L},{\bf R})}
\left(
{\bf L}_\varrho\circ c^{(1,0)}({\bf L}_\varrho,{\bf R}_\varrho)+
   {\bf R}_\varrho\circ c^{(0,1)}({\bf L}_\varrho,{\bf R}_\varrho)
\right)
(X)=-\frac{1}{2}\,X\,.
\end{eqnarray*}
ii) By the symmetry  it is sufficient to prove the assertion for
${\cal X}={\cal Y}$. From (\ref{defequ}) and i) we conclude
\begin{eqnarray*}
g(\Gamma({\cal X,X}),{\cal N})&=&
{\rm D}_ {\cal X}g\;({\cal X,N})-
\frac{1}{2}\,{\rm D}_ {\cal N}g\;({\cal X,X})=
{\rm D}_ {\cal X}g\;({\cal X,N})-g({\cal X},\Gamma({\cal X,N}))\\
&=&{\rm D}_ {\cal X}(g({\cal X,N}))
-g({\rm D}_ {\cal X}{\cal X},{\cal N})
-g({\cal X},{\rm D}_ {\cal X}{\cal N})
+\frac{1}{2}\,g({\cal X,X})\\
&=&
{\rm D}_ {\cal X}(\Tr{\cal X})-
\Tr {\rm D}_ {\cal X}{\cal X}
-g({\cal X,X})+\frac{1}{2}\,g({\cal X,X})=
-\frac{1}{2}\,g({\cal X,X})
\end{eqnarray*}
\newline
iii) From (\ref{DXN}) and i) we infer
$\nabla_{\cal X}{\cal N}={\rm D}_ {\cal X}{\cal N}
+\Gamma({\cal X},{\cal N})=
{\cal X}/2$
and the  last assertion follows since the torsion of
$\nabla$ vanishes.
$$
2{\cal R(X,Y,Z,N})=2{\cal R(Z,N,X,Y})
=2g\left({\cal Z,
\left(\nabla_X\nabla_Y-\nabla_Y\nabla_X-
\nabla_{[X,Y]_{\rm vf}}\right)N}\right)
=g\left({\cal Z},
{\cal\nabla_X Y-}{\cal \nabla_Y X-}{\cal \,[X,Y]_{\rm vf}}\right)=0
$$
{}\hfill $\Box$

\bigskip
Now, let $\cal X,Y$ be tangent vector fields  on
${\cal D}^1$ extended to some neighborhood  of
${\cal D}^1$ in ${\cal D}$, $g_\varrho({\cal X,N})=0$ for
$\varrho\in{\cal D}^1$.
Then ${}^1\!\nabla_{\!\cal X}{\cal Y}$
is the component of $\nabla_{\!\cal X}{\cal Y}$,
tangent to ${\cal D}^1$, \cite{Kob},
i.e.~
$${}^1\!\nabla_{\!\cal X}{\cal Y}=
\nabla_{\!\cal X}{\cal Y}-
g({\cal N},\nabla_{\!\cal X}{\cal Y}){\cal N}\,.
$$
Since ${\cal D}^1$ is a (locally) affine subspace of ${\cal D}$
the derivative ${\rm D}_{\!\cal X}{\cal Y}$ is tangent to
${\cal D}^1$  and we conclude from (\ref{Gamma})
and (\ref{l2})

\bigskip
\noindent
{\bf Proposition 3:}
{\sl
\begin{equation}
{}^1\!\nabla_{\!\cal X}{\cal Y}=
{\rm D}_{\!\cal X}{\cal Y}+
\Gamma ({\cal X,Y})+\frac{1}{2}\,g({\cal X,Y})\,{\cal N}\,,
\end{equation}
where $\Gamma $ is given by Proposition 1.}
{}\hfill$\Box$

\bigskip
Thus,  the component of $\nabla_{\cal\! X}{\cal Y}$
normal to ${\cal D}^1$
equals $-g({\cal X,Y}){\cal N}/2$ and  we obtain the curvature
tensor ${\cal R}^1$ of the submanifold ${\cal D}^1$
by the Gauss equation, \cite{Kob}:

\bigskip
\noindent
{\bf Proposition 4:}
\begin{equation}\llabel{R1}
{\cal R}^1({\cal X,Y,Z,W})={\cal R}({\cal X,Y,Z,W})+
\frac{1}{4}(
g({\cal X,Z})g({\cal Y,W})-g({\cal Y,Z})g({\cal X,W}))\,,
\end{equation}
{\sl where $\cal R$ is given by Proposition 3.}
\hfill$\Box$
\section{ scalar Curvature}
\subsection{General Theorem}

The scalar curvature at $\varrho\in{\cal D}^{(1)}$ we  denote by
$\cal S_\varrho$ resp.~${\cal S}_\varrho^1$.
It is given by
\begin{equation}\llabel{S}
{\cal S}_\varrho^{(1)}=
\sum_{X\neq Y\in {\cal B}^{(1)}}{\cal K}^{(1)}(X,Y)\,,
\end{equation}
where we sum up over pairs of different elements of  an orthogonal basis
${\cal B}^{(1)}$ of
${\rm T}_\varrho{\cal D}^{(1)}$.
The sectional curvature  of the plane generated  by  orthogonal
vectors $X$ and $Y$ equals  due to Propositions 2 and 4
\begin{equation}\llabel{K}
{\cal K}^{(1)}(X,Y)=
\frac{1}{g(X,X)g(Y,Y)}{\cal R}(X,Y,X,Y)
\quad+\quad\left(\;\frac{1}{4}\;\right)\,,
\end{equation}
where
\begin{eqnarray}
{\cal R}(X,Y,X,Y)&=&
g(\Gamma (X,Y),\Gamma (X,Y))-g(\Gamma (X,X),\Gamma (Y,Y))\nonumber\\
&&\llabel{prop2}
+
{{\rm D}}^2_{\!X,Y}g\;(X,Y)
-\frac{1}{2}
{{\rm D}}^2_{X,X}g\;(Y,Y)
-\frac{1}{2}{{\rm D}}^2_{Y,Y}g\;(X,X)\,.
\end{eqnarray}

It turns out that ${\cal S}^1$ and ${\cal S}$
differ at $\varrho\in {\cal D}^1$ only by a constant depending on $n$.
Indeed, let the basis ${\cal B}$ consists of the normal vector
${\cal N}_\varrho=\varrho$ and a certain basis  ${\cal B}^1$
of ${\rm T}_\varrho{\cal D}^1$.
Then by Lemma 1 the normal vector does not give any contribution
to $\cal S$ in (\ref{S}).  Since ${\cal S}^1$ is a sum of
$2\left({n^2-1\atop 2}\right)$ terms we get

\bigskip\noindent
{\bf Corollary 1}: {\sl Let $\varrho\in{\cal D}^1$. Then}
\begin{equation}
{\cal S}_\varrho^1={\cal
S}_\varrho+\frac{(n^2-1)(n^2-2)}{4}\,.
\end{equation}
{}\hfill$\Box$

Thus it is sufficient to consider the not normalized case.
In order to formulate our theorem we introduce the functions
$h_1,\dots,h_4$ complex analytic in every argument
in a certain neighborhood of ${\Bbb R}_+^{\;3}\subset{\Bbb C}^3$.
We define them for different arguments by
\begin{eqnarray}\llabel{h1}
h_1(x,y,z)&=&
\frac{c(x,y)-z\, c(x,z)\,c(y,z)}{(x-z)(y-z)c(x,z)c(y,z)}\\
\llabel{h2}
h_2(x,y,z)&=&
\frac{\left(c(x,z)-c(y,z)\right)^2}{(x-y)^2c(x,y)c(x,z)c(y,z)}\\
\llabel{h3}h_3(x,y,z)&=&z\,
\frac{(\ln c)'(z,x)-(\ln c)'(z,y)}{x-y}\\
\llabel{h4}h_4(x,y,z)&=&z\,(\ln c)'(z,x)\;(\ln c)'(z,y)\,,
\end{eqnarray}
otherwise  we go to the limit,
$h_1(x,x,z):=\lim_{y\rightarrow x}h_1(x,y,z)$ etc.
Indeed,
using the general properties (\ref{c0})-(\ref{c6})
of the Morozova-Chentsov
function $c$ one easily verifies, that all  these limits  exist.
By Riemann' theorem about removable singularities the resulting
functions  are in fact complex analytic in every argument. A similar
reasoning applies to other functions in this section  involving so called
divided differences.

For simplicity of notation
we consider the spectrum  $\sigma(\varrho)$
as an $n$-tuple with
possibly repeated elements if some eigenvalues appear with multiplicities,
but, nevertheless, we write $x\in\sigma(\varrho)$. Thus,
e.g.~$\sum_{x\neq y\in\sigma(\varrho)}term(x,y)$
always means a sum over $n^2-n$ pairs of eigenvalues, also if some or all
eigenvalues numerically coincide. Hence, this sum is equivalent to the more
extensive expression $\sum_{i\neq j=1,\dots,n}term(\lambda_i,\lambda_j)$.
A similar reasoning applies to  other sums with one or three indices of
summation. In the following summations the indices $x,y,z$ run through the
spectrum in this sense and  $i,j,k$ through $1,\dots,n$. This convention
enables us to formulate the main theorem of this section.

\bigskip\noindent
{\bf Theorem 1: }{\sl The scalar curvature on the manifold
$({\cal D},g)$ of $n\times n$ positive matrices  equals
\nopagebreak
\begin{eqnarray}\label{theorem}
{\cal S}_\varrho&=&
\sum_{x,y,z\in\sigma(\varrho)}h(x,y,z)-
\sum_{x\in\sigma(\varrho)}h(x,x,x)\,,\qquad \mbox{where}\\
\label{h}
h(x,y,z)&:=&h_1(x,y,z)-\frac{1}{2}h_2(x,y,z)+2h_3(x,y,z)-h_4(x,y,z)\,.
\end{eqnarray}}

\medskip\noindent
Before we prove the Theorem we give some conclusions and
remarks.

First of all, there is some ambiguity in the choice  of the function $h$.
Indeed, we can replace $h$ by any  function with
the same symmetrization not changing the  sum.
In particular we can replace $h$ by its symmetrization $\hs$,
\begin{equation}
\hs(x,y,z):=\frac{1}{3}\left(h(x,y,z)+h(y,z,x)+h(z,x,y)\right)\,.
\end{equation}
In the above Theorem
we tried to find a certain minimal formulation with different types
of divided differences, so that the involved functions have no
singularities if arguments coincide. However, for a concrete
function $c$ another
representation could be more convenient (see the examples below).

Going to the limit in (\ref{h1})-(\ref{h4})
results in
\begin{equation}\llabel{hxxx}
h(x,x,x)=
\frac{15}{8x}-3x^2\,c''(x,x)\,.
\end{equation}
Moreover, we see that (\ref{theorem}) is, actually, a trace formula.
Indeed,
we can look at (\ref{theorem}) as a  trace  of functions
of $\varrho$, that means
\begin{eqnarray*}
{\cal S}_\varrho&=&
\Tr\left\{\left(\Tr\left\{\left(\Tr h(s,t,\varrho)
\right)_{|t=\varrho}\right\}\right)_{|s=\varrho}\right\}
+\Tr\left\{-\frac{15}{8}\varrho^{-1}+
3\varrho^2c''(\varrho,\varrho)\right\}\,.
\end{eqnarray*}
Thus,  the scalar curvature can equal well be written as a threefold
complex integral around the
spectrum of $\varrho$.

\bigskip\noindent
{\bf Corollary 2:}
\begin{eqnarray}
{\cal S}_\varrho&=&
\oint\!\!\oint\!\!\oint
h(\xi,\eta,\tau)\,\Tr\frac{1}{\xi-\varrho}\,
\Tr\frac{1}{\eta-\varrho}\,\Tr\frac{1}{\tau-\varrho}\,
{\rm d}\xi\,{\rm d}\eta\,{\rm d}\tau
+\Tr\left\{-\frac{15}{8}\varrho^{-1}+
3\varrho^2c''(\varrho,\varrho)\right\}\,.
\end{eqnarray}
{}\hfill$\Box$

A further consequence concerns the curvature
at the most mixed state of ${\cal D}^1$. For
$\varrho=\frac{1}{n} \bbox{1}$ we obtain
$${\cal S}^1_\varrho=(n^3-n)\,h(1{/}n,1{/}n,1{/}n)+
\frac{(n^2-1)(n^2-2)}{4}$$
leading to

\bigskip\noindent
{\bf Corollary 3:} {\sl The scalar curvature on ${\cal D}^1$ at
$\varrho=\frac{1}{n}{\bbox{1}}$ equals
\begin{equation}\llabel{S{/}n}
{\cal S}^1_\varrho=
\frac{(n^2-1)(17n^3-4n-24\,c''(1/n,1/n))}{8n}\,.
\end{equation}
In particular, in the case of the smallest, largest and Kubo-Mori metric
related to the Morozova-Chentsov functions (\ref{MCf}) holds}
$$
{\cal S}^1_\varrho=
\frac{(n^2-1)(5n^2-4)}{8}\,,\quad
\frac{(1-n^2)(7n^2+4)}{8}\,\;\mbox{ resp. }
\frac{(n^2-1)(n^2-4)}{8}\,.
$$
{}\hfill$\Box$
\newline This confirms
the result obtained for the Kubo-Mori metric at
$\varrho=\frac{1}{n}{\bbox{1}}$ in \cite{Petz94}, Theorem 6.2.
For the minimal metric   this coincides with
Corollary 3 of \cite{Di99}.
The differing factor  is due to a factor $1/4$ in the
metric.

Finally, let ${\cal S}(\lambda_1,\dots,\lambda_n):=
{\cal S}_\varrho$, where $\lambda_i$ are the eigenvalues of $\varrho$.
Then, using $\hs$ instead of $h$, it is not difficult to verify, that the
following  recurrences  hold

\bigskip\noindent
{\bf Corollary 4:}
\vspace{-.5cm}
\begin{eqnarray}
(n-3)\,{\cal S}(\lambda_1,\dots,\lambda_n)&=&
\sum_i{\cal S}(\lambda_1,\dots,
{}^{\stackrel{\displaystyle{i}}{\vee}},
\dots,\lambda_n)-\sum_{i<j}{\cal S}(\lambda_i,\lambda_j)\,,
\\
{\cal S}(\lambda_1,\dots,\lambda_n)&=&
\sum_{i<j<k}{\cal S}(\lambda_i,\lambda_j,\lambda_k)-(n-3)\sum_{i<j}
{\cal S}(\lambda_i,\lambda_j)\,,\qquad n> 3
\end{eqnarray}
{}\hfill$\Box$

\bigskip
\noindent
{\bf Proof of Theorem 1:}
Since the metric is invariant under the
${\rm U}(n)$-conjugation on $\cal D$
we fix  a  diagonal $\varrho$ with the spectrum
$\sigma(\varrho):=\left\{\lambda_1,\dots,\lambda_n\right\}$.
We set
$${\rm b}_{ij}:={\rm
e}_{ij}+{\rm e}_{ji}\qquad\mbox{and}\qquad
\widetilde{\rm b}_{ij}:={\bf i\,}({\rm e}_{ij}-{\rm e}_{ji})\,,
$$
where ${\rm e}_{ij}$, $i,j=1,\dots,n$, are the standard matrices
with entries zero or one.
Then
$${\cal B}:=
\left\{{\rm b}_{ii}|1\leq i\leq n\right\}\cup
\left\{{\rm b}_{ij}|1\leq i<j\leq n\right\}\cup
\left\{\widetilde{\rm b}_{ij}|1\leq i<j\leq n\right\}
$$
is an orthogonal basis and in the sum (\ref{S}) there appear
terms of, essentially, three types, say A, B and C.
To see this, we first observe that the sectional curvature
vanishes for pairs of basis vectors with
disjoint  index sets, because ${\cal K}(b,b')$ is built,
finally, by products of diagonal resolvent
matrices and $b,b,b',b'$ (cf.~(\ref{nnmetric}) and (\ref{S2})).
By the same reason ${\cal
K}(b,b')$ depends only on the eigenvalues corresponding to
the indices involved. Moreover, using
the U$(n)$-invariance immediately follows
$$ {\cal K}({\rm b}_{ii},{\rm b}_{ij})=
 {\cal K}({\rm b}_{ii},\widetilde{\rm b}_{ij})
 \,\qquad
 {\cal K}({\rm b}_{ik},{\rm b}_{jk})=
 {\cal K}({\rm b}_{ik},\widetilde{\rm b}_{jk})=
 {\cal K}(\widetilde{\rm b}_{ik},\widetilde{\rm b}_{jk})\,\quad
\mbox{for }i\neq j\neq k\neq i\,, $$
because the arguments of $\cal K$ generate conjugated planes.
For example, in the case of  the second equation let
${\rm U}={\rm diag}(\dots,1,\dots,{\bf i},\dots,1,\dots )$
with $\bf i$ in the
$j$-th position.
Then ${\rm U}\varrho {\rm U}^*=\varrho$,
${\rm U}{\rm b}_{ik}{\rm U}^*={\rm b}_{ik}$ and
${\rm U}{\rm b}_{jk}{\rm U}^*=\widetilde{\rm b}_{jk}$.
The other identities follow by a similar choice of U.

Thus, all non vanishing terms in the sum (\ref{S}) are equal to one of the
expressions
${\cal K}({\rm b}_{ii},{\rm b}_{ij})$,
${\cal K}({\rm b}_{ij},\widetilde{\rm b}_{ij})$ or
${\cal K}({\rm b}_{ik},{\rm b}_{jk})$, which can be obtained from
${\cal K}({\rm b}_{11},{\rm b}_{12})$,
${\cal K}({\rm b}_{12},\widetilde{\rm b}_{12})$ and
${\cal K}({\rm b}_{13},{\rm b}_{23})$ by changing the indices at $\lambda$.
Hence, the knowledge of the dependence of these three terms on
$\lambda_1,\lambda_2$ resp.~$\lambda_1,\lambda_2,\lambda_3$ allows for
determining  (\ref{S}). Let this dependence be given by
the  three functions $A$, $B$ and $C$, that means
\begin{equation}\llabel{ABC}
A(x,y):={\cal K}({\rm b}_{11},{\rm b}_{12})\,,\quad
B(x,y):={\cal K}({\rm b}_{12},\widetilde{\rm b}_{12})\,,\quad
C(x,y,z):={\cal K}({\rm b}_{13},{\rm b}_{23})\,,
\end{equation}
where we assume on the right hand sides that the eigenvalues
$\lambda_1,\lambda_2,\lambda_3$ equal the  independent variables
$x,y,z$.
Of course, $B$ and $C$ are symmetric in $x,y$.

The computation of $A$, $B$ and $C$  requires many straightforward
calculations based on Propositions 1 and 2. We do not  present
them in full details and  explain only
some   essential intermediate results for $B$.
For a detailed verification the use of {\sl Mathematica}
or a similar  program is suggested. Moreover, the  function $A$ we can
obtain  from $C$ as we will see in Lemma 3.

\bigskip\noindent
{\bf Lemma 2:}\newline
\begin{eqnarray}\llabel{lemmaB}
B(x,y)&=&{{2 - (x+y)\,c(x,y)}\over
    {{{\left(x-y \right) }^2}\,c(x,y)}}
 -\frac{x\,(\ln c)'(x,y)^2}{4}
 -\frac{y\,(\ln c)'(y,x)^2}{4}
 -\frac{x\, (\ln c)'(x,y)-y\,(\ln c)'(y,x)}{x-y}\\
C(x,y,z)&=&\frac{1}{4}\left(
3h_1(x,y,z)-h_1(y,z,x)-h_1(z,x,y)\right)\nonumber\\
&+&\frac{1}{8}\left(
h_2(x,y,z)+h_2(y,z,x)+h_2(z,x,y)\right)-\frac{h_4(x,y,z)}{4}\nonumber\\
\llabel{lemmaC}&+&
\left\{\left(\frac{c(z,y)-c(x,y)}{2(x-z)^2c(x,z)\,c(y,z)}+
\frac{z\,(\ln c)'(z,y)}{2(x-z)}\right)
+\{x\leftrightarrow y\}\right\}
\end{eqnarray}

\bigskip\noindent
{\bf Proof:}
According to (\ref{ABC}) we  set $n=2$, $\lambda_1=x$, $\lambda_2=y$ and
$$
\varrho:=\left(\begin{array}{cc}x&0\\0&y\end{array}\right)\,,\quad
X:={\rm b}_{11}=2{\rm e}_{11}=
\left(\begin{array}{cc}2&0\\0&0\end{array}\right)\,,\quad
Y:={\rm b}_{12}={\rm e}_{12}+{\rm e}_{21}=
\left(\begin{array}{cc}0&1\\1&0\end{array}\right)\,;\quad
X,Y\in{\rm T}_\varrho{\cal D}\,.
$$
Equation (\ref{metric1}) yields
$g(X,X)=4c(x,x)=4/x$ and $g(Y,Y)=2c(x,y)$.
Further, using Proposition 1 we find
\begin{eqnarray*}
\Gamma(X,X)&=&
\frac{1}{(2\pi{\bf i})^2}\oint\!\!\oint
\frac{2c(\xi,\eta)}{(\xi-x)^2(\eta-x)}\,{\rm d}\xi\,{\rm d}\eta\;\,
c({\bf L},{\bf R})^{-1}(X)
=-\frac{1}{x}\,X\\
\Gamma(X,Y)&=&
\frac{1}{(2\pi{\bf i})^2}\oint\!\!\oint
\frac{c(\xi,\eta)}{(\xi-x)^2(\eta-y)}\,{\rm d}\xi\,{\rm d}\eta\;\,
c({\bf L},{\bf R})^{-1}(Y)
=(\ln c)'(x,y)\,Y\\
\Gamma(Y,Y)&=&
\frac{1}{(2\pi{\bf i})^2}\oint\!\!\oint
\left\{
\frac{2c(\xi,\eta)}{(\xi-x)(\xi-y)(\eta-x)}-
\frac{c(\xi,\eta)}{(\xi-x)^2(\eta-y)}\right\}
\,{\rm d}\xi\,{\rm d}\eta\;\,
c({\bf L},{\bf R})^{-1}({\rm e}_{11})\\
&+&
\frac{1}{(2\pi{\bf i})^2}\oint\!\!\oint
\left\{
\frac{2c(\xi,\eta)}{(\xi-x)(\xi-y)(\eta-y)}-
\frac{c(\xi,\eta)}{(\xi-y)^2(\eta-x)}\right\}
\,{\rm d}\xi\,{\rm d}\eta\;\,
c({\bf L},{\bf R})^{-1}({\rm e}_{22})\\
&&\\&=&
\left(
\frac{2(1-x \,c(x,y))}{x-y}-x \,c'(x,y)\right)\,{\rm e}_{11}
+\left(
\frac{2(1-y \,c(x,y))}{y-x}-y \,c'(y,x)\right)\,{\rm e}_{22}
\end{eqnarray*}
and, therefore,
\begin{eqnarray*}
g(\Gamma(X,Y),\Gamma(X,Y))&=&
\frac{2\,c'(x,y)^2}{c(x,y)}\\
-g(\Gamma(X,X),\Gamma(Y,Y))&=&
\frac{4(1-x\,c(x,y))}{x^2(x-y)}-\frac{2\,c'(x,y)}{x}\,.
\end{eqnarray*}
For the  second order derivatives  we obtain from (\ref{nnmetric})
\begin{eqnarray*}
-\frac{1}{2}{\rm D}_{\!X,X}g\;(Y,Y)&=&
-\frac{1}{(2\pi{\bf i})^2}\oint\!\!\oint
\frac{8\,c(\xi,\eta)}{(\xi-x)^3(\eta-y)}
\,{\rm d}\xi\,{\rm d}\eta=
-4\,c''(x,y)\\
{\rm D}_{\!X,Y}g\;(X,Y)-\frac{1}{2}{\rm D}_{Y,Y}g\;(X,X)
&=&\frac{1}{(2\pi{\bf i})^2}\oint\!\!\oint
\frac{8\,c(\xi,\eta)}{(\xi-x)^2(\eta-x)(\eta-y)}
\,{\rm d}\xi\,{\rm d}\eta=
\frac{4(1+2x^2c'(x,y))}{x^2(y-x)}
\end{eqnarray*}
Inserting all that into (\ref{K}), (\ref{prop2}) and (\ref{ABC})
finishes the proof of (\ref{lemmaB}).
The treatment of $C$ is  analogous, however it needs much  more
numerical effort.
\hfill$\Box$

\bigskip \noindent
{\bf Lemma 3:}
\begin{eqnarray}\llabel{lemma3a}{\rm i)}&&\qquad
A(x,y)=2C(x,y,x)\\
\llabel{lemma3b}
{\rm ii)}&&\qquad
B(x,y)=2C(x,x,y)+2C(y,y,x)
\end{eqnarray}

\bigskip\noindent
{\bf Proof:}
We set $n=3$, $\lambda_1=\lambda_3=x$, $\lambda_2=y$,
$X={\rm b}_{13}$ and
$Y={\rm b}_{23}$. Then
$g({\rm b}_{11},{\rm b}_{11})=
|\!|{\rm b}_{11}|\!|^2=|\!|{\rm b}_{33}|\!|^2=4/x$,
$|\!|X|\!|^2=2/x$,
$|\!|{\rm b}_{12}|\!|^2=|\!|Y|\!|^2=
2c(x,y)$ and
\begin{eqnarray*}
A(x,y)&=&{\cal K}_\varrho({\rm b}_{11},{\rm b}_{12})
=\frac{x}{8c(x,y)}\,{\cal R}
({\rm b}_{11},{\rm b}_{12},{\rm b}_{11},{\rm b}_{12})\,,\\
C(x,y,x)&=& {\cal K}_\varrho({\rm b}_{13},{\rm b}_{23})
=\frac{x}{4c(x,y)}\,{\cal R}
({\rm b}_{13},{\rm b}_{23},{\rm b}_{13},{\rm b}_{23})
={\cal K}_\varrho(X,Y)\,.
\end{eqnarray*}
Now we use the   ${\rm U}(3)$-symmetry of conjugation.
For this purpose let
$$u:=\left(\begin{array}{ccc}
\sqrt{2}/2&0&\sqrt{2}/2\\
0&1&0\\
\sqrt{2}/2&0&-\sqrt{2}/2
\end{array}\right)\in{\rm U}(3)\,.
$$
Then $u\varrho u^*=\varrho$ and
${\cal R}_\varrho(X,Y,X,Y)=
{\cal R}_\varrho(X',Y,X',Y')$ with
$
X':=uX u^*=\left({\rm b}_{11}-{\rm b}_{33}\right)/2
$ and
$
Y'=uY u^*=\sqrt{2}\left({\rm b}_{12}-{\rm b}_{23}\right)/2$.
Thus,
\begin{eqnarray*}
C(x,y,x)&=&{\cal K}(X,Y)=\frac{x}{4c(x,y)}{\cal R}_\varrho(X',Y',X',Y')=
\frac{x}{4c(x,y)}\frac{1}{8}{\cal R}
({\rm b}_{11}{-}{\rm b}_{33},{\rm b}_{12}-{\rm b}_{23} ,
 {\rm b}_{11}{-}{\rm b}_{33},{\rm b}_{12}-{\rm b}_{23})\\
&=&\frac{x}{32c(x,y)}
\big\{{\cal R}({\rm b}_{11},{\rm b}_{12},
               {\rm b}_{11},{\rm b}_{12})+
      {\cal R}({\rm b}_{33},{\rm b}_{23},
               {\rm b}_{33},{\rm b}_{23})+\dots\big\}\\
&=&\frac{x}{32c(x,y)}\big\{
|\!|{\rm b}_{11}|\!|^2\,|\!|{\rm b}_{12}|\!|^2\,A(x,y)+
{|\!|\rm b}_{33}|\!|^2\,|\!|{\rm b}_{23}|\!|^2\,A(x,y)
\big\}=\frac{1}{2}\,A(x,y)\,.
\end{eqnarray*}
All the terms indicated by dots in the above equation vanish.
From (\ref{Prop2}), (\ref{S2}) and
(\ref{nnmetric})  we see that these terms are built by products of
the four arguments of the curvature tensor and resolvents. Now,
considering the indices of the arguments of the neglected terms
it is not difficult to see, that these products
(e.g.~for ${\cal R}({\rm b}_{11},{\rm b}_{12},{\rm b}_{33},{\rm
b}_{12})$), or at
least their traces
(e.g.~${\cal R}({\rm b}_{11},{\rm b}_{12},
{\rm b}_{11},{\rm b}_{23})$),
vanish. This finishes the proof of (\ref{lemma3a}).

To prove (\ref{lemma3b}) we give  $C(x,x,y)$ resulting
from (\ref{lemmaC}). Then (\ref{lemma3b}) is immediately verified.
\begin{eqnarray}\llabel{Cxxy}
C(x,x,y)&=&\lim_{z\rightarrow x}C(x,z,y)\nonumber\\
&=&\frac{2 - (x+y)\,c(x,y)}{4\,(x-y )^2\,c(x,y)}
 +\frac{x\,(\ln c)'(x,y)^2}{8}
 -\frac{y\,(\ln c)'(y,x)^2}{4}
 -\frac{x\, (\ln c)'(x,y)-y\,(\ln c)'(y,x)}{4\,(x-y)}
\end{eqnarray}
{}\hfill$\Box$

Now we can proceed with the computation of $\cal S_\varrho$ starting with
(\ref{S}),
\begin{eqnarray*}
{\cal S}_\varrho&=&
4\sum_{i\neq j}
{\cal K}({\rm b}_{ii},{\rm b}_{ij})
+2\sum_{i\neq j}
{\cal K}({\rm b}_{ij},\widetilde{\rm b}_{ij})
+4\!\!\!\sum_{\textstyle {i<j\,;\,k<l\atop (i,j)\neq( k,l)} }
{\cal K}({\rm b}_{ij},{\rm b}_{kl})
\\
&=&
\sum_{x\neq y}\Big\{
4A(x,y)+B(x,y)\Big\}+
4\sum_{x\neq y\neq z\neq x}C(x,y,z)\\
&=&
4\sum_{x,y,z}C(x,y,z)
+\sum_{x}\Big\{
-4A(x,x)-B(x,x)+8C(x,x,x)\Big\}\\
&&+
\sum_{x,y}\Big\{
4A(x,y)+B(x,y)-4C(x,x,y)-8C(x,y,x)\Big\}\\
&=&
4\sum_{x,y,z}C(x,y,z)
-4\sum_{x}C(x,x,x)\,,
\end{eqnarray*}
where we used Lemma 3 and
$\sum_{x,y} 2C(x,x,y)=\sum_{x,y}\left(C(x,x,y)+C(y,y,x)\right)$.
Next, inserting  (\ref{lemmaC}) yields
\begin{eqnarray*}
{\cal S}_\varrho&=&
4\sum_{x,y,z}\left\{\frac{1}{4}\,h_1(x,y,z)+\frac{3}{8}\,h_2(x,y,z)
-\frac{1}{4}\,h_4(x,y,z)
+\frac{c(z,y)-c(x,y)}{(x-z)^2c(x,z)\,c(y,z)}+
\frac{z\,(\ln c)'(z,y)}{(x-z)}\right\}\\
&&-
4\sum_{x}C(x,x,x)\,.
\end{eqnarray*}
The  two fractions, whose sum has no singularities,
we replace by $-h_2(x,y,z)/2$ and $h_3(x,y,z)/2$, which have  the same
symmetrizations as the replaced terms. We end up with
$$
{\cal S}_\varrho=
\sum_{x,y,z}\left\{h_1(x,y,z)-\frac{1}{2}\,h_2(x,y,z)+2\,h_3(x,y,z)
-h_4(x,y,z)
\right\}
-4\sum_{x}C(x,x,x)
=\sum_{x,y,z}h(x,y,z)-\sum_{x}h(x,x,x)\,
$$
where we used $4C(x,x,x)=h(x,x,x)$, since $4C$ and $h$ have the same
symmetrization.
This finishes the proof of  Theorem 1.
\hfill$\Box$
\subsection{Examples}
In this paragraph we consider the scalar curvature for three
important examples of monotone metrics related to the Morozova-Chentsov
functions (\ref{MCf})  given in Section III.
\subsubsection{The smallest monotone  metric,
$c(x,y)=\frac{2}{x+y}$}
This metric was first introduced by Uhlmann in generalizing the
Berry phase to mixed states, \cite{Uh92b}.
Now it is often called Bures metric,
since it is, roughly speaking, the infinitesimal version of the
Bures distance of  mixed quantum states. It appears very natural in
the concept of purifications of mixed states and  intertwines between
the classical Fisher metric and the Study-Fubini metric on the complex
projective space representing pure states of a quantum system.
There are many papers dealing with this metric. Some references where
given in the Introduction,
for curvature results see \cite{Di93a,Di99}. Nevertheless we
explain this example, at least to see, that the whole machinery presented
here works.

Equations (\ref{h1})-(\ref{h4}) yield for $c(x,y)=\frac{2}{x+y}$
$$
h_1(x,y,z)=\frac{1}{2(x+y)}\,,\qquad
h_2(x,y,z)=\frac{x+y}{2(x+z)(y+z)}\,,\qquad
h_3(x,y,z)=h_4(x,y,z)=\frac{z}{(x+z)(y+z)}
$$
Thus, by Theorem 1  the scalar curvature
of $\cal D$ at $\varrho$ equals
\begin{eqnarray*}
{\cal S}_\varrho&=&\sum_{x,y,z}h(x,y,z)-
\sum_{x}h(x,x,x)=
\sum_{x,y,z}\left\{
\frac{1}{2(x+y)}-\frac{x+y}{4(x+z)(y+z)}+\frac{z}{(x+z)(y+z)}\right\}
-\frac{3}{8}\sum_x\frac{1}{x}\\
&=&
\sum_{x,y,z}\left\{
\frac{1}{4(x+z)}+\frac{1}{4(y+z)}-\frac{x+y}{4(x+z)(y+z)}+
\frac{z}{(x+z)(y+z)}\right\}
-\frac{3}{8}\sum_x\frac{1}{x}
\end{eqnarray*}
Thus $$ {\cal S}_\varrho=
\frac{3}{2}\sum_{x,y,z}\frac{z}{(x+z)(y+z)}-
-\frac{3}{8}\sum_x\frac{1}{x}\,, $$ where we sum up over the
spectrum of $\varrho$. This confirms the result obtained in
\cite{Di99}, where one can find also expressions for the scalar
curvature in terms of invariants of $\varrho$. The different
factor is due to an other normalization of the metric, as already
mentioned above. In particular, we get from Corollary 1 for $n=2$
and $\sigma(\varrho)=\{\lambda_1,\lambda_2\}$,
$\lambda_1+\lambda_2=1$, $${\cal
S}_\varrho^1=\frac{9}{2(\lambda_1+\lambda_2)}+\frac{3}{2}=6\,$$
i.e.~${\cal D}^1$ is a space of constant curvature for this
exceptional dimension, In our normalization  it looks locally like
a sphere of radius 2. This was one of the first results concerning
this metric, \cite{Uh92b}. For higher dimensions a similar result
does not hold, ${\cal D}^{(1)}$ is not a locally symmetric space,
\cite{Di93a}.
\subsubsection{The largest monotone metric,
$c(x,y)=\frac{x+y}{2\,x\,y}$}
This metric  belongs to the series of monotone metrics
related to
$$c_s(x,y):=\frac{x^{2\alpha} +y^{2\alpha}}{2(x\,y)^{\alpha+1/2}}\,,
\qquad0\leq\alpha\leq\frac{1}{2}\,,$$
see \cite{Petz}, also for further references of its application.
It is the  monotone metric  obtained in the simplest way besides
the Hilbert-Schmidt metric.
Indeed,
$$g_\varrho(X,Y)=
\frac{1}{2}\Tr X\varrho^{-1}(\varrho \,Y+Y\varrho)\varrho^{-1}=
\frac{1}{2}\Tr \varrho^{-1}(X Y+Y X)$$
and further  quantities one  can get without
using the integral representations, e.g.
\begin{eqnarray*}
{\rm D}_{\! Z}g\;(X,Y)&=&
-\frac{1}{2} \Tr\varrho^{-1}Z\varrho^{-1}(X Y+Y X)\,,\qquad
{\rm D}^2_{\!Z,W}g\;(X,Y)=
\frac{1}{2} \Tr\varrho^{-1}(W\varrho^{-1}Z+Z\varrho^{-1}W)\varrho^{-1}
(XY+YX)\\
\Gamma(X,Y)&=&
\frac{1}{2({\bf L}+{\bf R})}(XY+YX)-
\frac{1}{2}(X\varrho^{-1}Y+Y\varrho^{-1}X)\,.
\end{eqnarray*}
Thus, using Proposition 2,
 it is easy to write down the curvature tensor.
For the scalar curvature we rely on our Theorem 1,
\begin{eqnarray*}
h_1(x,y,z)&=&h_3(x,y,z)=\frac{-z}{(x+z)(y+z)}\,,\qquad
h_2(x,y,z)=\frac{2z^2}{(x+y)(x+z)(y+z)}\,,\\
h_4(x,y,z)&=&\frac{xy}{z(x+z)(y+z)}\,,\qquad
h(x,x,x)=\frac{-9}{8x}\,,
\end{eqnarray*}
and  proceed with
\begin{eqnarray*}
{\cal S}_\varrho&=&
\sum_{x,y,z}(h_1+2h_3)(x,y,z)-\frac{1}{2}\sum_{x,y,z}h_2(x,y,z)
-\sum_{x,y,z}h_4(x,y,z)-\sum_x h(x,x,x)\\
&=&
-3\sum_{x,y,z}\frac{z}{(x+z)(y+z)}
-\sum_{x,y,z}\left(\frac{x}{(x+z)(y+z)}-\frac{z}{2(x+z)(y+z)}\right)
-\sum_{x,y,z}\frac{x\,y}{z(x+z)(y+z)}+\sum_x\frac{9}{8x}\\
&=&
-\frac{5}{2}\sum_z z\,\left(\sum_x\frac{1}{x+z}\right)^2
-\sum_z\left(\sum_x\frac{x}{x+z}\right)\left(\sum_x\frac{1}{x+z}\right)
-\sum_z\frac{1}{z}\left(\sum_x\frac{x}{x+z}\right)^2+
\sum_x\frac{9}{8x}\\
&=&
-\frac{5}{2}\sum_z z\,\left(\sum_x\frac{1}{x+z}\right)^2
+n\sum_{x,z}\frac{1}{x+z}+\left(\frac{9}{8}-n^2\right)\sum_x\frac{1}{x}\,.
\end{eqnarray*}
Inserting
$$\sum_x\frac{1}{x+t}=
\Tr \frac{1}{\varrho+t}=-\frac{\chi'(-t)}{\chi(-t)}
\,,$$
where $\chi(t):=\sum e_{n-i}(-t)^i$ is the
characteristic polynomial of $\varrho$, we get  the scalar
curvature as a trace of a  function of $\varrho$ (diagonal or not):
\begin{equation}
{\cal S}_\varrho=
-\frac{5}{2}\Tr \varrho\,\frac{\chi'(-\varrho)^2}{\chi(-\varrho)^2}
-n\, \Tr \frac{\chi'(-\varrho)}{\chi(-\varrho)}
+\left(\frac{9}{8}-n^2\right)\Tr \varrho^{-1}\,.
\end{equation}
Moreover, we can express the scalar curvature in terms of the elementary
invariants $e_i$. For this purpose we consider the matrix
$${\cal E}_{ij}:=\left\{\begin{array}{lll}
1&\quad&\mbox{for }i+1=j\\
(-1)^{n-j}e_{n+1-j}&&\mbox{for }i=n\\
0&&\mbox{otherwise}
\end{array}
\right.
$$
It has the same characteristic polynomial as $\varrho$ and, therefore,
it is  conjugated (not unitarily) to $\varrho$ in the generic case of
different eigenvalues. But the set of $\varrho$ with different eigenvalues
is dense in $\cal D$ and by continuity of the scalar curvature we conclude

\bigskip
\noindent
{\bf Proposition 5:}
\begin{equation}
{\cal S}_\varrho=
-\frac{5}{2}\Tr {\cal E}\,\frac{\chi'(-{\cal E})^2}{\chi(-{\cal E})^2}
-n\, \Tr \frac{\chi'(-{\cal E})}{\chi(-{\cal E})}
+\left(\frac{9}{8}-n^2\right)\Tr {\cal E}^{-1}\,.
\end{equation}
\hfill$\Box$

\noindent A similar result was obtained in \cite{Di99} for the previous
example.

\subsubsection{The Kubo-Mori metric,
$c(x,y)=\frac{\log(x)-\log(y)}{x-y}$} This metric was considered
by Petz, Hiai and Toth, \cite{Petz94,Petz}, where they pointed out
its importance in information theory and quantum statistics and
obtained partial results concerning the sectional and scalar
curvature. In \cite{Petz94} the scalar curvature at the trace
state, cf.~Corollary 3, and the sectional curvature
$K_\varrho(X,Y)$  where found for $X,Y$ commuting  with $\varrho$.
Our Theorem 1 leads to a closed expression for the scalar
curvature at arbitrary  $\varrho$ in terms of its eigenvalues.
This result is the basis for the last section, where we consider
the conjecture mentioned in the Introduction.

For the  functions $h_i$ we obtain
\begin{eqnarray*}
h_1(x,y,z)&=&h_3(x,y,z)=
\frac{\displaystyle
\frac{z-x+z\,(\ln x-\ln z)}{(x-z)(\ln x-\ln z)}-
\frac{z-y+z\,(\ln y-\ln z)}{(y-z)(\ln y-\ln z)}}{x-y}\\
h_2(x,y,z)&=&h_1(x,y,z)+h_1(y,z,x)+h_1(z,x,y)\\
h_4(x,y,z)&=&
\frac{(z-x+z\,(\ln x-\ln z))\;(z-y+z\,(\ln y-\ln z))}{
z\,(x-z)(\ln x-\ln z)\;\;\;\;(y-z)(\ln y-\ln z)\;\;}\,,\quad
h(x,x,x)=\frac{-1}{8x}\,.
\end{eqnarray*}
We observe, that now $h_2$ is already a symmetric function. The typical
term seen in these equations is
$$ \frac{z-x+z\,(\ln x-\ln z)}{(x-z)(\ln x-\ln z)}=
z\,(\ln c)'(z,x)
$$
and we set
\begin{eqnarray}\llabel{d}
d(x,y,z)&:=&\frac{3}{2}h_1(x,y,z)-h_4(x,y,z)=
z\,\left(3\frac{(\ln c)'(z,x)-(\ln c)'(z,y)}{2(x-y)}
-(\ln c)'(z,x)\,(\ln c)'(z,y)\right)\nonumber\\
&=&
3\;
\frac{\displaystyle
\frac{z-x+z\,(\ln x-\ln z)}{(x-z)(\ln x-\ln z)}-
\frac{z-y+z\,(\ln y-\ln z)}{(y-z)(\ln y-\ln z)}}{2(x-y)}
-\frac{(z-x+z\,(\ln x-\ln z))\;(z-y+z\,(\ln y-\ln z))}{
z\,(x-z)(\ln x-\ln z)\;\;\;\;(y-z)(\ln y-\ln z)\;\;}.
\end{eqnarray}
If two arguments  coincide the values of $d$ are  easily found, we
do not list them. Clearly, $d$ has the same symmetrization as $h$
and $h(x,x,x)=d(x,x,x)$. Thus we infer from Corollary 1 and
Theorem 1

\bigskip
\noindent {\bf Theorem 2: }{\sl The scalar curvature  of the
Kubo-Mori metric on ${\cal D}^{(1)}$  at $\varrho$ equals}
\begin{equation}
{\cal S}_\varrho^{(1)}= \sum_{x,y,z\in\sigma(\varrho)}d(x,y,z)-
\sum_{x\in\sigma(\varrho)}d(x,x,x)\quad
\left(\;+\frac{(n^2-1)(n^2-2)}{4}\;\right)\,.
\end{equation}
\hfill$\Box$

\bigskip
\noindent

The following graph  of ${\cal S}^1$ restricted to the (open)
2-dimensional simplex ($n=3$) $$\left\{{\rm
diag}(\lambda_1,\lambda_2,\lambda_3)\,|
\,\lambda_1+\lambda_2+\lambda_3=1\\ \,,\lambda_i>0\right\}$$ of
diagonal densities will be instructive for  the aim of the next
section.
\begin{figure}
\centerline{\epsfig{file=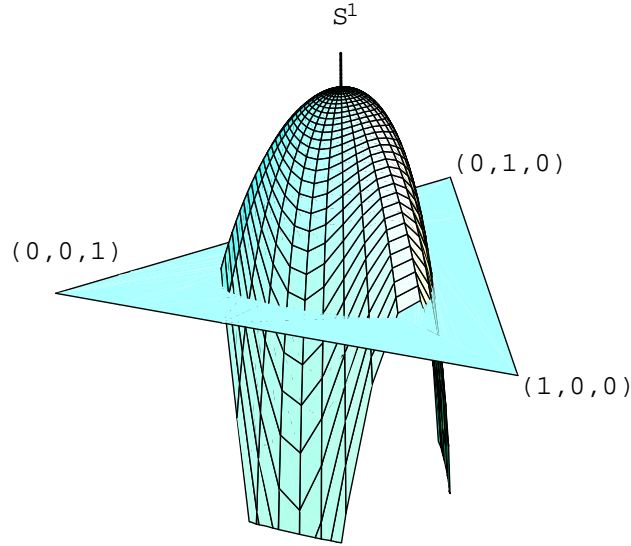,height=12cm}}
\caption{Graph of ${\cal S}^1$ for $n=3$}
\end{figure}
Moreover, using this Theorem one confirms  all numerical examples
given for $n=3$ in \cite{Petz94}, however one has to work with
high numerical precision if two eigenvalues are close to each
other.

It should be mentioned that   formula (\ref{Prop2}) for the
curvature   simplifies for this metric, because the terms with
second order derivatives of $g$ cancel out. Due to  general
properties of quadrilinear forms (see \cite{Kob}) this is
equivalent to the vanishing of the sum of the corresponding three
terms in (\ref{prop2}). We do not explain this in detail and refer
also to  \cite{Petz94}, formula (6.4). In our approach  one could
proceed as follows: Under the additional assumption
$$(y-x)c(y,x)+(x-z)c(x,z)+(z-y)c(z,y)=0$$ for a general
Morozova-Chentsov function c one  can show $${\rm
D}_Xg\;(X,Y)={\rm D}_Yg\;(X,X)\,.$$ Clearly, this assumption is
fulfilled in this example. Now it is not difficult to conclude
$$2{{\rm D}}^2_{\!X,Y}g\;(X,Y) -{{\rm D}}^2_{X,X}g\;(Y,Y) -{{\rm
D}}^2_{Y,Y}g\;(X,X)=0\,,$$ and, therefore, (\ref{prop2}) reads
\begin{equation}
{\cal R}(X,Y,X,Y)= g(\Gamma (X,Y),\Gamma (X,X))-g(\Gamma
(X,X),\Gamma (Y,Y))\,.
\end{equation}
\section{Monotonicity of ${\cal S}^1$ under
mixing for the Kubo-Mori metric} The manifold ${\cal D}^1$
represents the space of faithful mixed states of a $n$-dimensional
quantum system and carries the following partial order (for
details and further related facts used here we refer e.g.~to
\cite{AlUh}): $\rho$ is called more mixed than $\varrho$,
$\rho\mmix \varrho$, if there exists a trace preserving stochastic
map $\alpha$ such that $\rho=\alpha(\varrho)$. If $\lambda_i,
\mu_i$  are the eigenvalues of $\rho$ and $\varrho$ in decreasing
order this relation becomes  equivalent to
$$\lambda_1+\dots+\lambda_k\leq \mu_1+\dots+\mu_k,\qquad
k=1,\dots,n\,.$$ Petz conjectured that the scalar curvature of the
Kubo-Mori metric on ${\cal D}^1$ behaves monotonously under this
partial order, more precisely:

\bigskip
\noindent{\bf Conjecture,} \cite{Petz94,Petz96} {\bf:}
\vspace{-0.5cm}
\begin{equation}
\rho\mmix \varrho\quad\Longrightarrow\quad {\cal
S}^1_{\rho}\geq{\cal S}^1_{\varrho}
\end{equation}
An immediate consequence would be that the scalar curvature on
${\cal D}^1$ attains its  maximum  at the most mixed state,
i.e.~at the trace state. Hence, $\max {\cal S}^1={\cal
S}^1_{\bbox{1}/n}=(n^2-1)(n^2-4)/8$, see Corollary 3.

In this section we shows this important conjecture  up to the
concavity of a certain function in three variables, respectively
up to some weaker consequences of this concavity. For this purpose
we use the results for the Kubo-Mori metric obtained in the
example before.

Since the scalar curvature is invariant under  U$(n)$ conjugation,
it is sufficient to consider the conjecture on the
$(n-1)$-dimensional  (open) simplex $\left\{{\rm
diag}(\lambda_1,\dots,\lambda_n)|\sum\lambda_i=1\,,
\lambda_i>0\right\}$. Clearly, FIG.~1 supports the above
hypothesis in the case $n=3$.

To deal with a general $n$ we consider the symmetrization $\hs$ of
the function $h$, resp.~$d$, see (\ref{d}). A straightforward
calculation yields
\begin{eqnarray}
&&\hspace{-1cm}
\hs(x,y,z)=\frac{1}{3}(d(x,y,z)+d(y,z,x)+d(z,x,y))
          =\frac{1}{3}(h(x,y,z)+h(y,z,x)+h(z,x,y))\nonumber\\
&=&\frac{1}{6}h_2(x,y,z)+
\frac{1}{3}\left(h_2(x,y,z)-h_4(x,y,z)-h_4(y,z,x)-h_4(z,x,y)\right)
\nonumber\\&&\nonumber\\ &=& \frac{(y-z)^2 (\ln x-\ln y) (\ln
x-\ln z) -
   (x-z)^2 (\ln x-\ln y) (\ln y-\ln z) +
   (x-y)^2 (\ln x-\ln z) (\ln y-\ln z)}
{6\,(x-y)(x-z)(y-z)(\ln x-\ln y)(\ln x-\ln z)(\ln y-\ln z)
}\nonumber\\ &&\nonumber\\ \llabel{ds}&&+ \frac{ - x\,y\,( \ln x -
\ln y ) + x\,z\,( \ln x - \ln z ) - y\,z\,( \ln y - \ln z ) }
{3\,x\,y\,z\,( \ln x - \ln y )( \ln x - \ln z )( \ln y - \ln z )}
\end{eqnarray}
We make the following

\bigskip\noindent{\bf Assertion: }
{\sl The function $\hs$ is concave on ${\Bbb R}_+^3$.}

\bigskip
\noindent We do not have a formal proof of this assertion, but
some millions  of numerical tests confirmed
$$(1-t)\hs\left(P_1\right)+
t\,\hs\left(P_2\right)\leq \hs\left((1-t)P_1+t\,P_2\right)\,,
\qquad 0<t<1\,.$$
Moreover,  concavity is equivalent to
negative semi-definiteness of
$$\left( \frac{\partial^2\hs}{\partial
x_i\partial x_j}\right)_{i,j=1,2,3} $$
and the plots of its main minors
$$M_k(x,y,z):=
\det\left( \frac{\partial^2\hs}{\partial
x_i\partial x_j}(x,y,z)\right)_{1\leq i,j\leq k}\,,\quad k=1,2,3, $$
see Fig.~2-4,
suggest the correct alternating signs
we need. Since  $\hs$ is homogeneous  of degree -1,
the second order derivatives are homogeneous of degree
$-3$ and we can fix one coordinate
to 1 resulting in the   3D-plots
given below. Thus, the Assertion is certainly true
and we believe that a more or less difficult formal proof
will be found later.

What we really need to prove the conjecture are some weaker
properties of $\hs$. To formulate them let  $\hsd$ denote
 the first order partial  derivative with
respect to the first variable as is the previous sections.

\bigskip\noindent
{\bf Lemma 4:} {\sl If  the above Assertion  is true then for all
$x,y,\lambda,\mu\in{\Bbb R}_+$ with $x< y$ holds}
\begin{eqnarray}\llabel{lemma41}
0&\leq& 2\hsd(x,x,y)-\hsd(y,x,x)-2\hsd(y,x,y)+\hsd(x,y,y) \,,\\
0&\leq& \hsd(x,x,\lambda)-\hsd(y,y,\lambda) \,,\\ 0&\leq&
\hsd(x,y,\lambda)-\hsd(y,x,\lambda) \,,\\ 0&\leq&
\hsd(x,\lambda,\mu)-\hsd(y,\lambda,\mu) \,.
\end{eqnarray}

\noindent{\bf Proof:} If $\hs$ is concave then $[0,1]\ni
t\longmapsto\hs(\gamma(t))$, where
$\gamma(t):=(1-t)\,P_0+t\,P_1\,,$ is concave for all
$P_0,P_1\in{\Bbb R}_+^3$ and, in particular,
$$\frac{\rm d}{{\rm d}t}\hs(\gamma(t))_{\restriction_{t=0}}
\geq
\frac{\rm d}{{\rm d}t}\hs(\gamma(t))_{\restriction_{t=1}}\,.$$
Now the
inequalities follow if we set
\begin{eqnarray*}
P_0&:=&(x,x,y)\,,\quad (x,x,\lambda)\,,\quad  (x,y,\lambda)\;
\mbox{ resp. }(x,\lambda,\mu) \\ P_1&:=&(y,y,x)\,,\quad
(y,y,\lambda)\,,\quad (y,x,\lambda)\; \mbox{ resp.
}(y,\lambda,\mu))
\end{eqnarray*}
{}\hfill$\Box$

\bigskip\noindent
{\bf Theorem 3 :} {\sl If  the  Assertion about concavity of $\hs$
is true then the Conjecture about the monotonicity of the scalar
curvature is true.}

\bigskip
\noindent{\bf Proof:} Let $\rho\mmix\varrho$. Then there exists a
sequence $\varrho_i$, $i=1,\dots ,m$ such that
$$\rho=\varrho_m\mmix\varrho_{m-1}\mmix\dots\mmix\varrho_2\mmix
\varrho_1=\varrho$$ and the spectra of each consecutive pair
differ only in two eigenvalues. This is a often used standard
fact, for a proof see e.g.~\cite{AlUh}. Therefore, and by the
unitary invariance of ${\cal S}^1$, we may assume that $$
\rho=\mbox{diag}(x',y',\lambda_3,\dots,\lambda_n)\,,\qquad\
\varrho=\mbox{diag}(x,y,\lambda_3,\dots,\lambda_n)\, \quad
\mbox{with }x<y\,. $$ There are no further order assumptions for
the eigenvalues. Thus, the more mixed pair $(x',y')$ must be of
the form $$ \left(\begin{array}{c}x'\\y'\end{array}\right)=
\left(\begin{array}{c}x_t\\y_t\end{array}\right):=
\left(\begin{array}{cc}1-t&t\\t&1-t\end{array}\right)
\left(\begin{array}{c}x\\y\end{array}\right) $$ for a certain
$t\in[0,1]$ and it is sufficient to prove that
$t\mapsto{\cal
S}^1(x_t,y_t,\lambda_3,\dots,\lambda_n)
$
is nondecreasing at
$t=0$, i.e.~that $$\frac{{\rm d}}{{\rm d}t}{\cal
S}^1(x_t,y_t,\lambda_3,\dots,\lambda_n)_ {\restriction_{t=0}}=
(y-x)\left( \frac{\partial}{\partial x}-\frac{\partial}{\partial
y} \right) {\cal S}^1(x,y,\lambda_3,\dots,\lambda_n)\geq 0\,. $$
By Theorem 2 we have
\begin{eqnarray*}
\frac{1}{3}\,{\cal S}^1(x,y,\lambda_3,\dots,\lambda_n)&=&
\hs(x,x,y)+\hs(y,y,x)\\ &&+\sum_{i=3}^n
\left(\hs(x,x,\lambda_i)+\hs(y,y,\lambda_i)+2\hs(x,y,\lambda_i)\right)
+\sum_{i,j=3}^n
\left(\hs(x,\lambda_i,\lambda_j)+\hs(y,\lambda_i,\lambda_j)\right)\\
&&+ \dots\quad\mbox{(terms not depending on $x,y$)}
\end{eqnarray*}
and, therefore,
\begin{eqnarray*}
\frac{1}{3}\left( \frac{\partial}{\partial
x}-\frac{\partial}{\partial y} \right) {\cal
S}^1(x,y,\lambda_3,\dots,\lambda_n)&=&
2\hsd(x,x,y)-\hsd(y,x,x)-2\hsd(y,x,y)+\hsd(x,y,y)\\
&&+2\sum_{i=3}^n \left(\hsd(x,x,\lambda_i)-\hsd(y,y,\lambda_i) +
\hsd(x,y,\lambda_i)-\hsd(y,x,\lambda_i)\right)\\ &&+\sum_{i,j=3}^n
\left(\hsd(x,\lambda_i,\lambda_j)-\hs(y,\lambda_i,\lambda_j)\right)
\geq 0\,,
\end{eqnarray*}
where we used Lemma 4. {}\hfill$\Box$
\begin{figure}
\centerline{\epsfig{file=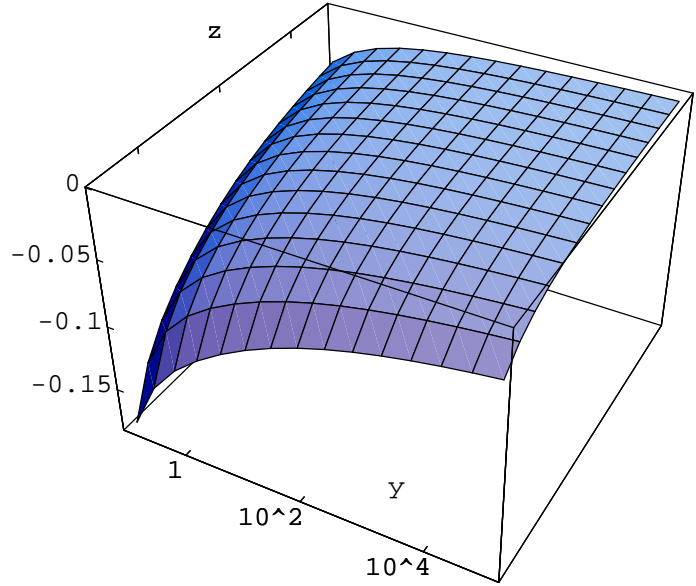,height=12cm}}
\caption{Graph of $M_1(1,y,z)$}
\end{figure}
\begin{figure}
\centerline{\epsfig{file=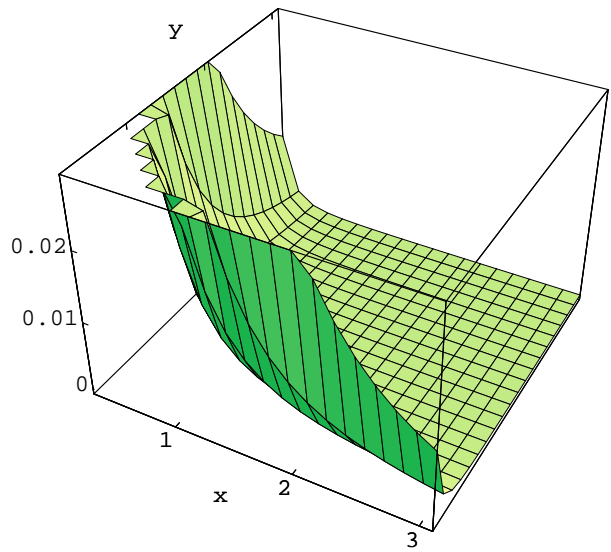,height=10.5cm}}
\caption{Graph of the second main minor $M_2(x,y,1)$}
\end{figure}
\begin{figure}
\centerline{\epsfig{file=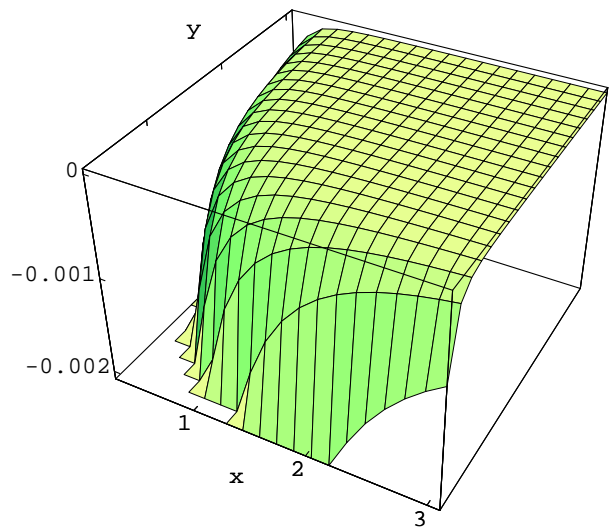,height=10.5cm}}
\nopagebreak
\caption{Graph of $M_3(x,y,1)$}
\end{figure}
\acknowledgments I would
like to thank  A.~Uhlmann (Leipzig) for enlighten and valuable remarks.
Moreover I am grateful to D.~Petz (Budapest) for stimulating discussions
during a stay at the Banach Center in Warsaw.

\end{document}